\documentstyle[epsf]{europhys}
\def\And{{\rm and\ }}

\def\stars{\bigskip\centerline{***}\medskip}

\newif\ifboo \boofalse

\input epsf
\begin{document}

\euro{}{}{}{}
\Date{December 1999}
\shorttitle{GEOMETRICAL PICTURE FOR SPIN GLASSES}
\title{A geometrical picture for finite dimensional spin glasses}
\author{J. Houdayer \And O. C. Martin}
\institute{
Laboratoire de Physique Th\'eorique et Mod\`eles Statistiques,\\
Universit\'e Paris-Sud, F-91405 Orsay,
France.\\
%\email{houdayer@ipno.in2p3.fr, martino@ipno.in2p3.fr}
}
%\rec{}{}

%\offprints{O.C. Martin}
%\mail{O.C. Martin}


\maketitle

\begin{abstract}
A controversial issue in spin glass theory is whether mean field
correctly describes $3$-dimensional spin glasses.
If it does, how can replica symmetry breaking
arise in terms of spin clusters in
{\it Euclidean} space?
Here we argue that there exist system-size
low energy excitations that are ``sponge-like'',
generating multiple valleys separated by diverging energy
barriers. The droplet model should be valid for length scales smaller
than the size of the system ($\theta > 0$), but nevertheless
there can be system-size excitations of constant energy without
destroying the spin glass phase. The picture we propose then combines
droplet-like behavior at finite length scales
with a potentially mean field behavior at the system-size scale.
\end{abstract}

\section{Introduction}
%\label{SectIntro}
The solution of the Sherrington-Kirkpatrick mean field
model of spin glasses shows that
its equilibrium states are organised in a hierarchy
associated with continuous replica symmetry breaking
(RSB)~\cite{MezardParisi87b}.
A working paradigm for some years has been that this type of
replica symmetry breaking also
occurs in finite dimensional spin glasses above the
lower critical dimension ($2 < d_l < 3$); we will call this school
of thought the {\it mean field picture}. The question
of whether this paradigm is correct is still the subject of
an active debate (see~\cite{MarinariParisi99b} and references therein).

The mean field hierarchical organisation of states
corresponds to valleys within valleys ...
within valleys. Though such a structure is appealing to many,
it seems to us necessary to describe how
it can possibly arise for spins
lying in {\it Euclidean} space. As an example,
consider the many nearly degenerate ground states
predicted by mean field; what is the nature
of the clusters of spins that flip when going from one
such state to another? It is not clear {\it a priori} that
mean field has much predictive power here for the
following reason. In any finite dimension,
there are clusters whose surface to volume
ratio is arbitrarily small. However this kind of object does
{\it not} arise in models without geometry such as the
(infinite range) Sherrington-Kirkpatrick model or mean field
diluted models (such as the Viana-Bray model); any cluster in those
models has a surface growing essentially as fast as its volume.
This key difference is very important in
spin glass models having up-down symmetry: when flipping
a cluster, the change in energy comes from the
surface only, but the change in quantities like the
overlap goes as the volume of the cluster.
Another reason for insisting on finding a geometrical
(sometimes called ``real space'') picture is that
the dynamics of a real spin glass is local in space, leading to
coherence effects that build up from small to large
length scales.
Realistic theories of spin glasses should
then allow for these scales by incorporating
the Euclidean geometry in which the spins are embedded.

Spin clusters in Euclidean space have generated much interest
since the early
80's~\cite{McMillan83} and have been studied in detail by Fisher and
Huse~\cite{FisherHuse86,FisherHuse88}.
These authors focused on understanding the properties of low lying
excitations above the ground state, and their objects of study
were localised compact clusters of spins (droplets).
However, their goal was not to come up with a
picture compatible with RSB; on the contrary,
their conclusions are in direct conflict with the
mean field picture. Unfortunately, there have been very few other
works based on geometrical points of view. The
possibility that droplets may be fractal rather than compact
has been considered several times~\cite{Villain86,OcioHammann90}
and has gained renewed interest
in the last few years~\cite{BouchaudDean95,JonasonVincent98}.
In our work, we want to deepen
the geometrical point of view and provide a coherent picture
of valleys in the energy landscape of finite dimensional spin glasses.
(All of the discussion that follows concerns energies; the
temperature is zero.)

We hope to convince the reader that
appropriately constructed clusters of spins that are neither
compact nor localised (i) have low energies,
and (ii) are separated by energy barriers that diverge in the
thermodynamic limit. These clusters occur on the
size of the whole system which is assumed to be finite but
arbitrarily large, so we call them
system-size clusters. From the ``spongy'' nature of these
clusters we can see how
the spin glass stiffness exponent $\theta$ can be
positive in spite of the possible presence of
system-size excitations of energy $O(1)$.
Within our picture,
the droplet model is expected to be valid at finite length scales
(corresponding to properties within one valley) while
a different picture, possibly mean-field-like, may be valid at
the scale of the whole system. (The two scales of validity of
course do not overlap.) We will provide
several plausibility arguments for this kind of
a mixed picture. No numerical evidence will be
presented here, rather we will
ask the reader to imagine what happens when one searches
for low lying states; we believe that this point of
view can uncover the essential qualitative properties of
the energy landscape.

\section{$\theta$ exponents in the droplet and scaling pictures}
For definiteness, we consider the Edwards-Anderson model
on an $L \times L \times L$ cubic lattice,
but our reasoning can be applied to any short range
spin glass model in dimension $d \ge 3$.
The Hamiltonian
has nearest neighbor couplings and no external field:
\begin{equation}
H = - \sum_{<ij>} J_{ij} S_i S_j
\end{equation}
The couplings $J_{ij}$ are independent
random variables with a symmetric distribution about $0$.
For simplicity, we take the
$J_{ij}$ to be continuous random variables so that generically
the ground state is non-degenerate in any finite volume.
In all that follows, we assume the ground state to be given
and investigate the low energy excitations above the ground state.

Fisher and Huse~\cite{FisherHuse86} define a droplet as
a cluster having a characteristic size, containing
a given site, and having the lowest possible energy if it is flipped.
Because of the locality of the
Hamiltonian, it is enough to consider droplets that are
connected in the sense that their spins are within the range
of the interaction couplings. Fisher and Huse
then argue that droplets
of characteristic radius $r$ have energies that
scale as $r^{\theta}$. Implicitly, their
construction considers $r$ given
while the lattice size $L \to \infty$; only after having reached the infinite
volume limit does one let $r$ grow. It is also possible to
consider $L$ and $r$ going to infinity {\it simultaneously}. This
is precisely what is done when one measures the exponent
$\theta$ by comparing periodic and anti-periodic boundary
conditions~\cite{McMillan85,BrayMoore86}. This way of imposing
boundary conditions introduces a {\it domain wall}
that splits the system into a ``left'' and a
``right'' part; one expects the
characteristic energy of such a domain wall to scale as $L^\theta$,
matching onto the
droplet scaling law. Clearly it is best to distinguish these two
methods and to keep in mind that there are two exponents:
the first, $\theta_l$
($l$ for local), associated with scales much
smaller than the size of the system; the second, $\theta_{dw}$,
associated with the introduction of a domain wall that splits
the system into a left and a right part.

Is $\theta_l=\theta_{dw}$? One might fear that a global
object such as a domain wall spanning the width of the
system will be sensitive to the shape of the box used
or to the way one introduces the domain wall~\cite{vanEnter90}.
But Fisher and Huse~\cite{FisherHuse88} have argued that
droplets should have similar local surface properties to those of an
interface formed with the periodic-anti-periodic boundary conditions.
In that case, one expects the droplet scaling
$E(r) \approx r^{\theta_l}$ to
extend to sizes $r \approx L$, leading to
$\theta_l=\theta_{dw}$. But one need not and should not conclude
that the lowest energy excitations on the length scale of $L$ obey
this energy scaling if their nature is {\it different} from domain walls.
Our claim is that in
spin glasses there do exist other kinds of excitations on the scale
of the whole system
that do not resemble domain walls but that
instead resemble sponges
(they have a non-zero surface to volume ratio
and are topologically highly non-trivial).
We expect these excitations to have energies smaller than
$O(L^{\theta_l})$. Then extrapolating the droplet picture
from finite size droplets to excitations spanning the whole
system misses some of the most important physics of
low lying states. It is thus necessary
to distinguish the local exponent
$\theta_l$ that describes
finite size excitations at $L = \infty$ from a
global exponent, hereafter called $\theta_g$, that
describes the energy scaling of the lowest energy
system-size excitations.
Mathematically, this simply means that the
limit $L$, $r \to \infty$ depends on the ratio $r/L$.
We have previously presented evidence for this kind of
dependence~\cite{HoudayerMartin98}
in a quite different
disordered system, the minimum matching problem.
That model has droplet-like excitations, allowing
one to introduce an exponent $\theta_l$ which satisfies
$\theta_l \ge 0$; one can also numerically measure $\theta_g$
and one finds $\theta_g \approx -0.5$.
In the rest of this article, we will
argue that this type of subtlety for global
properties also arises in spin glasses
so that the distinction between $\theta_l$
and $\theta_g$ is crucial.
Note that, as pointed
out by Fisher and Huse, $\theta_l > 0$
is a {\it necessary} condition for the existence of a
spin glass phase at finite temperature; however, no such condition applies
to $\theta_g$. In fact, within the mean field picture,
there are system-size excitations of constant energy, so
$\theta_g = 0$; the presence of such excitations does
{\it not} preclude a spin glass phase.

\section{Compact versus non-compact droplets}
Within the Fisher-Huse picture, the surface of a droplet
is rough in analogy with that of an interface in
a disordered medium.
Such a behavior is found for the surfaces created by
comparing periodic and anti-periodic boundary conditions.
The standard picture is thus that droplets are compact
({\it i.e.}, their surface to volume ratio goes to zero
as their volume grows) with
a rough surface having small overhangs and handles.
(To be specific, we identify the surface of a cluster
with the set of {\it bonds} connecting the cluster to its complement.)

A second scenario has
been suggested~\cite{BouchaudDean95,JonasonVincent98},
namely that droplets are not compact but fractal.
Here, we would like to consider a third scenario. It seems to
us that there may be a fixed length scale
below which droplets are compact, but beyond which
their surface to volume ratio no longer decreases because
handles proliferate and penetrate into their insides. Such objects
can best be thought of as sponges:
these are homogeneous at a coarse-grained level,
have handles everywhere, and are multi-connected.
Because of this property, the scenario we present here
can be relevant only in dimensions $d \ge 3$.
The essential feature of sponges is
that {\it both} their surface and their volume grow proportionally to their
``diameter'' to the power $d$.
This is to be contrasted with what happens in the two other scenarios.
In the case of compact Fisher-Huse droplets, the surface grows
as a power which is smaller than $d$. In the case of
fractal droplets, it is the volume that grows as a smaller power.

In both the fractal and sponge scenarios,
$\theta_{dw} \ne \theta_l$ and $\theta_{dw} \ne \theta_g$
because domain walls cannot mimick fractals or sponges.
However, in all three scenarios
(compact, fractal, or spongy droplets),
we are still left with the main question of this paper,
namely whether $\theta_g=\theta_l$. In what follows,
we shall present our arguments for $\theta_g \ne \theta_l$.
If necessary, the reader may assume that droplets
are compact, but in fact any scenario for droplets
can be assumed without affecting our arguments.

\section{Motivating $\theta_g \ne \theta_l$}
We begin with a class of finite size clusters that are not
of minimal energy (not droplets); they are only
of {\it relatively} low energy. To make these clusters tangible,
we think of constructing them
by a stochastic algorithm. The details of such an algorithm are
not an important issue here. Rather our goal is to understand
by simple reasonings the qualitative topological features
of the clusters that will be generated.

The algorithm starts with the ground state configuration of
spins. Suppose a seed site is given from which we build a
connected cluster by stochastically adding neighboring sites which seem
promising. For specificity,
one can imagine that one adds the
sites to the cluster {\it irreversibly} and in a near-greedy fashion
where only spins leading to the lowest instantaneous energies are
likely to be iteratively added.
What kind of clusters emerge from this construction?
Certainly these clusters must have a high connectivity, meaning
that for each site of the cluster, a significant fraction of
its neighbors on the cubic lattice are also part of the cluster.
This property should hold if the distribution of $J_{ij}$
is not too broad because generally
one has to flip more than one neighbor
of a spin to change the sign of its local field. A low energy
cluster will thus avoid stringy or one-dimensional parts.
(If on the contrary the distribution of
the $J_{ij}$ is broad, the clusters look rather like
percolating clusters~\cite{NewmanStein96c}.)
Furthermore, since a majority of the bonds are satisfied in the
ground state (recall that the ground state energy density is negative),
nearby sites will tend to all
join (or all {\it not} join) the cluster. In other
words, the algorithm leads to a ``cohesive'' length
${\ell}_c$ ($c$ for cohesion) below which the clusters are
compact. Beyond the length scale ${\ell}_c$,
the growth process might follow that of the
Eden model, leading to rough but compact clusters.
However, it is possible to have a different scenario where
beyond ${\ell}_c$ the
interface of the growing cluster is unstable
to branching and to the proliferation
of handles. The resulting clusters then probably
resemble sponges, with both a cluster and its
complement being multi-connected. One may define a posteriori
the scale ${\ell}_c$ for such an object
as the inverse of its surface to volume ratio,
or more physically from the peak in the
sponges' structure factor. Qualitatively,
${\ell}_c$ can be considered to be the
typical diameter a sphere of spins can grow to before
it hits the surface of the sponge.

The characteristic energies of these clusters
should grow significantly faster than that of droplets.
However, we expect most of the excess
energy of a cluster to come from the ``active'' sites that will
see their environment change if the growth process is
continued; these sites have extra unsatisfied
bonds that will become satisfied later on during the growth.
We shall thus think of the moving surface
or ``outside'' boundary of the cluster as concentrating most of
the excess energy.
(Note that by construction our growth rule is irreversible).

Overall, we then view our cluster growth process
as generating an expanding sponge with a multitude of handles beyond
the length scale ${\ell}_c$, leaving behind a low energy ``inside'';
if the algorithm is smart enough, there is reason to
believe that the energy
{\it density} of this inside will be the same as that of the ground state.

Now we want to extrapolate from finite size clusters
to system-size clusters.
We continue to grow our
sponge-like clusters but now we take into account the fact that
we work on a finite $L \times L \times L$ lattice.
As one increases the size of a sponge, its
``outside'' surface increases, but only until the sponge reaches
a diameter of about $L$. After that, the number of bonds
which are unsatisfied will decrease because
one can push the sponge's
outside surface out of the system. Suppose for instance
that one has free
boundary conditions. When the growing interface reaches
the edge of the lattice, it ``falls off'': there are no bonds
there to connect the inside of the cluster to its complement.
(For a trivial example, consider the compact cluster which
will grow to flip {\it all} the spins of
the lattice; its energy goes from
$L^{\theta_l}$ to zero as its interface goes to the edge of the system.)
In the case of periodic boundary conditions, the same phenomenon
occurs if the cluster touches itself by wrap-around.
We argued above that most of a sponge's
energy comes from its outside surface;
if that component of the surface is eliminated, one obtains a cluster
of small characteristic energy.
%The growth algorithm can create a huge number
%of such system-size clusters. (For this to hold it is enough
%for the algorithm to not be completely greedy
%in the choices for adding sites at each step. One then realises
%that getting the {\it best} cluster possible via this
%algorithm requires making the right
%choices a large number of times, a nearly impossible task; thus
%our gedanken algorithm is more of conceptual than
%practical use except perhaps for quite small systems.)
%The typical energy of the clusters generated by the
%algorithm may be large, but we
%are interested in the very lowest energies as in a
%random energy model. To be precise,
%the quantity of interest is the scale (as a function
%of $L$) of the {\it lowest} energy
%of such a system-size cluster. Presumably this is the same
%as the scale of the second or third lowest energy of this
%kind of excitation.

Given this gedanken construction, we now suggest
that the lowest excitations that flip a finite fraction of the
$L \times L \times L$ spins may be spongy and span the whole system;
we shall refer to them as {\it system-size} clusters.
Appealing to scaling arguments, their energies should
scale as $L^{\theta_g}$. The new nature of these excitations
(if they are sponges reaching the edge of
the system) suggests that $\theta_g \ne \theta_l$.

\section{Structure of the energy landscape}
To go from the ground state to one of these low-lying
sponge states, one has to grow its corresponding
cluster; during the growth, the
cluster will reach a maximum outside surface
where its characteristic size is $O(L)$.
Following the energy analysis for droplets,
the associated {\it energy barrier} between the ground state
and a system-size cluster should thus be at least $O(L^{\theta_l})$ and
may be much more.
This argument can be extended to the barriers
separating these different low energy configurations from
one another too. Our conclusion is
then that the Edwards-Anderson (EA) spin glass has an
energy landscape with many ``valleys'' separated
by diverging energy barriers as $L \to \infty$. {\it The cluster of spins
one must flip to go from one of our valleys to another
is sponge-like: both it and its complement span the whole
lattice.} The valleys
should be similar microscopically, that is any finite size window
will not permit one to distinguish the true ground state
from one of these excited states.
The reason is that ground states
in spin glasses are sensitive to changes
in the coupling constants; if we perturb a finite fraction of
the $J_{ij}$s even far away,
the ground state should change chaotically. Then the insides
of the low lying sponges are equally good
candidates for the new ground state, and all should be
statistically similar; in particular, they should have the same
energy density in the large volume limit.
At finite temperature, one can expect these valleys to give rise to a
multitude of states (if $\theta_g >0$, it is more appropriate
to call them metastable states). Because of the diverging
energy barriers, it will be nearly impossible
to equilibrate any lattice of
size $L \times L \times L$ as soon as $L$ is large
enough, and domain walls will ``freeze-in'' even
if their density is not particularly low.

\section{RSB with sponges}
One of the key questions that
discriminates between the droplet/scaling picture
and the mean field picture is the value
of $\theta_g$: it is implicitly assumed in the scaling
picture that $\theta_g = \theta_l > 0$, whereas in the
mean field picture, $\theta_g = 0$.
{\it If} we assume that there exists
constant energy sponge-like excitations and that
$\theta_g = 0$, we get
a ``real space'' picture of the EA spin glass
which is compatible with RSB. Indeed, with
$\theta_g = 0$, sponge excitations give rise
to a non-trivial overlap probability distribution. One can also look
at {\it window} overlaps~\cite{MarinariParisi98b}: for windows widths
larger than the characteristic size of the handles in the sponges,
one should see a non-trivial overlap distribution there too. Similarly,
one can look at the {\it energy} overlap probability distribution;
since the sponges are expected to have a finite surface to
volume ratio in the large $L$ limit, this kind of overlap
also gives a non-trivial signal of RSB.
Finally, {\it within} each valley, the exponent $\theta_l$
continues to describe the size dependence of a droplet's energy and
thus determines the decrease of the spin-spin correlation
function.

\section{Periodic-anti-periodic boundary conditions revisited}
%\label{SubSectionTheta}
If there are sponge-like
system-size excitations, their characteristic energies
cannot be measured by comparing
periodic to anti-periodic boundary conditions. The reason is
that such boundary conditions do not couple to
topologically non-trivial excitations. As a result, even if
$\theta_{dw} > 0$, one cannot exclude the presence of system-size
excitations of constant energy, {\it i.e.}, $\theta_g=0$.
Of course, in the case of the EA model in {\it two} dimensions,
there are no handles or sponges, so it is possible
that the droplet picture gives a correct description
of the system on {\it all} scales and that $\theta_g = \theta_l$.

To probe topologically unconstrained
low energy excitations in three dimensional spin glasses, one
has to develop other methods. The main work in this
direction is due to Palassini and
Young~\cite{PalassiniYoung99a,PalassiniYoung99b}.
They consider the effect on the ground state of randomly changing
some of the bonds at the boundaries. Their data
for window overlaps far from these boundaries show a power
law behavior in $L$, tending towards a trivial overlap function
in the large $L$ limit. If this extrapolation is correct, then
the mean field picture is wrong.
To sustain the mean field picture, one would have to
claim that the power law behavior seen by Palassini and Young
is only a transient,
and that for ``sufficiently'' large $L$ the non-trivial overlaps
no longer decrease but survive with a finite (non-zero) amplitude.
Only in this case, with $\theta_g =0$, can one expect
sponge-like excitations to appear upon changing a finite
number of bonds. Indeed, if $\theta_g > 0$, {\it droplet} excitations
couple to the changes in bond values, but sponges will not as $L \to \infty$.

There is also a direct way to test numerically for the
necessity to go beyond periodic vs. anti-periodic
boundary conditions. Suppose one can
extract the lowest energy excited state
subject to the constraint of having
an overlap with the ground state that is smaller than some
fixed value (say 0.5).
Our first claim is that the characteristic scale of this
excitation energy is $L^{\theta_g}$ with $\theta_g < \theta_l$. Our second
claim is that the cluster of spins associated with the
excitation will have a sponge-like non-trivial topology.
The main obstacle impeding this kind of
numerical test of our picture is the problem of {\it finding} excited states.
The only truely reliable way to find these states is by
a branch and bound algorithm as was done for the
minimum matching problem~\cite{HoudayerMartin98}. At present,
branch and bound algorithms for spin glasses cannot
treat systems larger than $L=5$, so it is important to improve
significantly the algorithmic tools currently available.

\section{Discussion and conclusions}
%\label{SectConclusions}
To arrive at our picture, we used a cluster growth gedanken
algorithm. It is possible to use other ways
to guess at the nature of system-size excitations.
Imagine searching for the lowest energy
excitation that flips a finite fraction of the spins on the lattice.
There will be an interface separating the spins that are flipped
from those that are not. An initial guess is that the
interface corresponds to a rough domain wall. This is what
happens if there is a non-zero surface tension. For our spin glass
system where in the ground state a majority of the bonds are
satisfied, a {\it typical} surface indeed has a positive surface tension.
But optimised domain wall energies (as measured for instance
using periodic vs. anti-periodic boundary conditions) teach us that
the {\it effective} surface tension of an interface
is zero if it is allowed to distort and take advantage of
the unsatisfied bonds in the ground state. (Note that in
our system a surface energy
can have a negative value as long as it does not separate
an inside from an outside.) With a zero surface tension, clearly
the interface separating the flipped and not flipped spins
will take advantage of topological freedoms,
so handles will permeate through the whole system.
This leads directly to a picture of sponge-like excitations.
There is an analogue of this in
systems without disorder, e.g., in
microemulsions of two fluids whose interfacial tension is zero.
This leads to
bi-continuous phases (see \cite{Cates90} and
references therein), structures
that are often called sponges.
%see also gyroidal phases~\cite{LuzzatiGulik68}.
%If such systems are probed using too simple boundary
%conditions (analogous to periodic-anti-periodic), the
%multi-connected phases do not appear.

Yet another way to guess at the nature of system-size excitations is
obtained if we {\it require} mean field to be relevant in finite
dimensions. Since excitations in mean field
models cannot have surface to volume ratios going to zero in
the large volume limit,
we are forced to consider excitations
in finite dimensional spin glasses which also have this property,
and this leads one quite directly to sponge structures.

We hope to have convinced the reader that sponge-like system-size
excitations are likely
to be relevant structures in the energy landscape of
short range spin glasses. The picture that then follows
is one where the droplet model is probably valid
at any fixed length scale whereas a different picture
(perhaps mean-field-like) gives the correct description of
{\it system-size} excitations.
It would be useful to deepen our geometrical picture in
several ways. An important problem is to justify
quantitatively $\theta_g < \theta_l$. Furthermore, taking
mean field as a guide,
under what conditions might the lowest valleys
associated with sponges be equilibrium states with
excess (free) energies of $O(1)$ rather than metastable states with
diverging excess (free) energies?

\stars
We are indebted to J.-P. Bouchaud for extremely stimulating
discussions. We also thank our colleagues
I. Campbell, H. Hilhorst, and E. Vincent for constructive criticisms.
J.H. (presently at Institut fur Physik, J. Gutenberg Universitat,
Mainz) acknowledges a fellowship from
the MENESR for this work, and O.C.M. acknowledges support 
from the Institut Universitaire
de France. The LPTMS is an Unit\'e Mixte de Recherche
de l'Universit\'e Paris~XI associ\'ee au CNRS.

\bibliographystyle{prsty}
\bibliography{../references}

\end{document}